\documentclass[twocolumn,showpacs,preprintnumbers,amsmath,amssymb,aps]{revtex4}
\input epsf
\usepackage{graphicx}
\begin{document}

\title{Point defect dynamics in two-dimensional colloidal crystals} 
\author{A. Lib{\' a}l$^{1,2}$, C. Reichhardt$^{1}$, and
C.J. Olson Reichhardt$^{1}$}
\affiliation{ 
{$^1$}Center for Nonlinear Studies and Theoretical 
Division, 
Los Alamos National Laboratory, Los Alamos, New Mexico 87545\\
{$^2$}Department of Physics, University of Notre Dame, Notre Dame, 
Indiana 46556
}

\date{\today}
\begin{abstract}
We study the topological configurations and dynamics of individual point 
defect vacancies and interstitials in a 
two-dimensional crystal of colloids interacting via a repulsive
Yukawa potential.
Our Brownian dynamics simulations show that the diffusion mechanism 
for vacancy defects 
occurs in two phases. The defect can glide along the crystal
lattice directions, and it can rotate during an excited topological 
transition configuration to assume a different direction
for the next period of gliding. 
The results for the vacancy defects 
are in good agreement with recent experiments. 
For interstitial point defects, which were not studied in the experiments, 
we find several of the same modes of motion as in the
vacancy defect case along
with two additional 
diffusion pathways. The interstitial defects are more mobile than 
the vacancy defects due to the more two-dimensional nature of the
diffusion of the interstitial defects.
\end{abstract}
\maketitle

\vskip2pc

\section{Introduction}
Topological defects in two-dimensional crystals
are relevant to a number of condensed matter systems including 
vortices in type-II superconductors \cite{Frey}, 
Wigner crystals \cite{Fisher}, magnetic bubble arrays \cite{Westervelt},   
atoms on surfaces, and dusty plasmas \cite{Goree}.
The creation of topological defects such as dislocations and disclinations 
leads to a two step melting process and   
an intermediate hexatic phase \cite{Nelson,Standburg,Maret}.  
Topological defects play a role in the mechanical response of the 
system, such as when a shear is applied to the crystal, and also
determine how effectively the crystal can be pinned to a disordered
substrate.
More recently, there has been growing interest in studying topological 
defects in two-dimensional
systems on curved surfaces, such as colloidal particles on 
the surface of a liquid drop \cite{Bausch,Bausch2}. 
In addition to providing insight into how defects can affect 
equilibrium and nonequilibrium properties of these systems, 
understanding how individual topological defects move 
would also be valuable for technological applications,
such as the nanoscale or mesoscale engineering of 
new two-dimensional materials. 

In an effort to understand the dynamics of individual point 
defects in two dimensions, recent experiments were
conducted in which point defects were artificially created in a 
two-dimensional 
colloidal suspension by manipulating the colloids with 
optical tweezers \cite{Pertsinidis,Pertsinidis2,Ling}.
Here, a defect is created by removing one or two colloids
from a perfect triangular
lattice, resulting in a mono- or divacancy.  The defect configurations
were shown to have a lower symmetry than the triangular lattice,
and specific topological arrangements were
characterized by the arrangement of fivefold and sevenfold coordinated
colloids around the cores of the mono- and divacancies.  
The experiments showed that the trajectories of both types of defects followed
the major axis directions of the triangular colloidal crystal. 
As a result, at short times the 
defect diffusion had one-dimensional characteristics.

In this work, we examine the topological configurations and dynamics of
single point vacancies and interstitial colloids
in a two-dimensional triangular colloidal crystal. 
Only vacancies were considered
in the experimental work \cite{Pertsinidis,Pertsinidis2,Ling}, 
but here we compare the behavior of vacancies to that of interstitial 
defects.
We find that the point defects have numerous stable topological 
configurations as well as several frequently appearing 
excited configurations.
The diffusive thermal motion of the defects is aligned with the major
axes of the crystal, as also seen in the experiments.
As the defect moves, it switches between gliding configurations,
which contain two fivefold coordinated particles, and transient 
configurations, which contain three or more fivefold coordinated particles.
The defect is able to change its gliding orientation in the transient
configurations.  
The mobility of interstitial defects is greater than that of vacancy
defects, and we show that this is a result of the fact that
interstitial defects are more likely to undergo reorientations in 
their gliding
direction than vacancy defects, giving the diffusion of the interstitial
defects a more two-dimensional character.
We discuss the implications of our work for other systems, 
such as the effect of the motion
of interstitial and vacancy defects 
on transport in two-dimensional vortex lattices. 

\section{Simulation methods}

We simulate a two dimensional colloidal crystal composed
of $N=1116\pm 1$ particles using Brownian dynamics. 
The system size is $L_x=31a_0$ and $L_y=18\sqrt{3}a_0$, where
distances are measured in units of the colloid lattice constant $a_0$, and
where we
employ periodic boundary conditions in both the $x$ and the $y$ directions. 
The overdamped equation of motion for an individual colloid $i$ is 
\begin{equation}
\eta \frac{d{\bf r}_i}{dt} = {\bf f}_i = {\bf f}_{cc} + {\bf f}_i^T 
\end{equation}
where the damping coefficient $\eta=1$ in simulation units.
Here the colloid-colloid interaction force is 
${\bf f}_{cc} = -\sum_{j\neq i}^{N} \nabla_i U(r_{ij}){\bf {\hat{r}}}_{ij}$, 
where $r_{ij}=|{\bf r}_{i}-{\bf r}_{j}|$ 
is the distance between particles located at ${\bf r}_{i}$ and ${\bf r}_{j}$,
and ${\bf {\hat r}}_{ij}=({\bf r}_{i}-{\bf r}_{j})/r_{ij}$.
We represent the colloid interaction via a Coulomb potential that is 
screened by the presence of ions in the liquid phase, giving us the 
Yukawa form $U(r_{ij}) =  E_0 q^2 \exp(-\kappa r_{ij})/r_{ij}$. 
The unit of energy is $E_0=Z^{*2}/(4\pi\epsilon\epsilon_0 a_0)$ where
$Z^*$ is the unit of charge, $\epsilon$ is the solvent dielectric constant, 
$q=1$ is the dimensionless charge on each colloid, and
$1/\kappa=0.16 a_0$ is the screening length.
Our system is in the strongly charged low volume fraction limit, allowing
us to neglect hydrodynamic interactions.
The thermal force ${\bf f}_i^T$ 
is introduced as Langevin random kicks that obey the 
equations $\langle {\bf f}_i^T(t)\rangle =0$ and 
$\langle {\bf f}_i^T(t){\bf f}_j^T(t^\prime)\rangle = 
2\eta k_B T \delta_{ij}\delta(t-t^\prime)$. 
Temperatures are reported with respect to the melting
temperature $f_T^m$ for the two-dimensional crystal.
Throughout this work, we consider temperatures 
in the range 
$0.37 \le f_T/f_T^m \le 0.62$, 
which is high enough to produce
observable defect motion, but low enough to fall below the 
temperature at which a proliferation of fivefold and
sevenfold defect excitations begins to occur.
After initializing the system in a triangular lattice, we either add
or subtract a single colloid from a location near the center of the
sample.
We then measure the time evolution of the system over a long time period
of $2 \times 10^8$ simulation time steps. 

The periodic boundary conditions in our sample prevent the individual
defects from annihilating.  Since the interaction between topological
defects is well known to be long ranged, finite size effects are always
a concern in a periodic system.  This is particularly important in 
studies of multiple interacting defects.
In our case, we are working with only
a single defect, which experiences a perfectly symmetrical interaction
with its images across the periodic boundaries.  This tends to minimize
the impact of the periodic boundaries.  To check for finite size effects,
we tested larger systems and did not observe any changes in the diffusive
behavior.  Additionally, our results for the vacancy defects are in 
good agreement with experiments, offering further evidence that 
we have avoided finite size effects.

\section{Point defect Configurations}

To characterize the defect configuration and identify the position of the
defect, we use a
Voronoi cell construction performed after the system
has relaxed into a stationary state from its initial configuration. 
We can identify the coordination number $z_i$ of each colloid by counting
the sides of the polygons in the Voronoi construction.
In an ideal triangular lattice, all particles are sixfold coordinated
with $z_i=6$. 
Our system has either a missing particle or an extra particle, and the 
defect core is surrounded by
a number of $z_i \ne 6$ particles.  
We identify distinct topological configurations of the defect
based on the Voronoi cell picture, adopting the same 
notation as used in Refs.~\cite{Pertsinidis2,Nelson2}. 
All of the possible configurations that exist in this system 
conserve the average sixfold coordination of the particles, 
\begin{equation} 
\sum_{i=1}^N z_i = 6N .
\end{equation}

\begin{figure}
\includegraphics[width=3.5in]{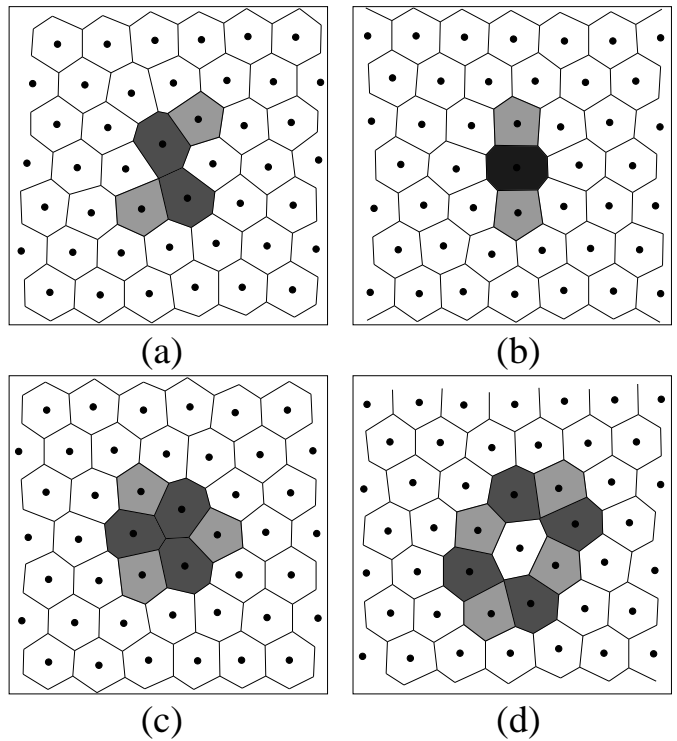}
\caption{Voronoi cell construction for 
commonly observed vacancy defect configurations. Colloid positions
are indicated by dots.  The Voronoi cells are colored according to the
coordination number $z_i$ of the colloids: white, $z_i=6$; light grey, 
$z_i=7$; dark grey, $z_i=5$; very dark grey, $z_i=8$. 
Only a $6 a_0 \times 6 a_0$ portion of the full system is shown.
(a) A twofold crushed vacancy ($V_{2a}$). 
(b) A split vacancy ($SV$) centered on a $z_i=8$ colloid. 
(c) A threefold symmetric vacancy ($V_3$). 
(d) A fourfold symmetric excited configuration $V^\prime_4$ containing
eight colloids with $z_i \ne 6$.
}
\label{fig:defecttypes}
\end{figure}

In Fig.~\ref{fig:defecttypes}
we illustrate the most prevalent configurations for the vacancy defect.
Fig.~\ref{fig:defecttypes}(a) shows that
the twofold crushed configuration ($V_{2a}$) consists of two nearly parallel 
pairs of $z_i=5$ and $z_i=7$ particles.
The split configuration ($SV$) for the vacancy, illustrated in 
Fig.~\ref{fig:defecttypes}(b), contains three particles with
$z_i \ne 6$, forming an almost straight line with 
two $z_i=5$ particles on opposite sides of a
$z_i=8$ particle. 
The third equilibrium configuration for the vacancy, 
the threefold symmetric configuration ($V_3$) shown
in Fig.~\ref{fig:defecttypes}(c),
consists of three pairs of particles with $z_i=5$ and $z_i=7$
arranged around the outside of a triangle that is centered on the vacancy.
These configurations are the same as those observed
for the monovacancy in the colloidal experiments \cite{Pertsinidis2,Ling}. 
Figure \ref{fig:defecttypes}(d) illustrates an important excited
configuration that we find for the vacancy.  This fourfold symmetric
configuration, termed $V^\prime_4$, contains eight colloids with $z_i \ne 6$.
It is short lived and
plays a role in the mobility of the vacancy, as will be described
below.

\begin{figure}
\includegraphics[width=3.5in]{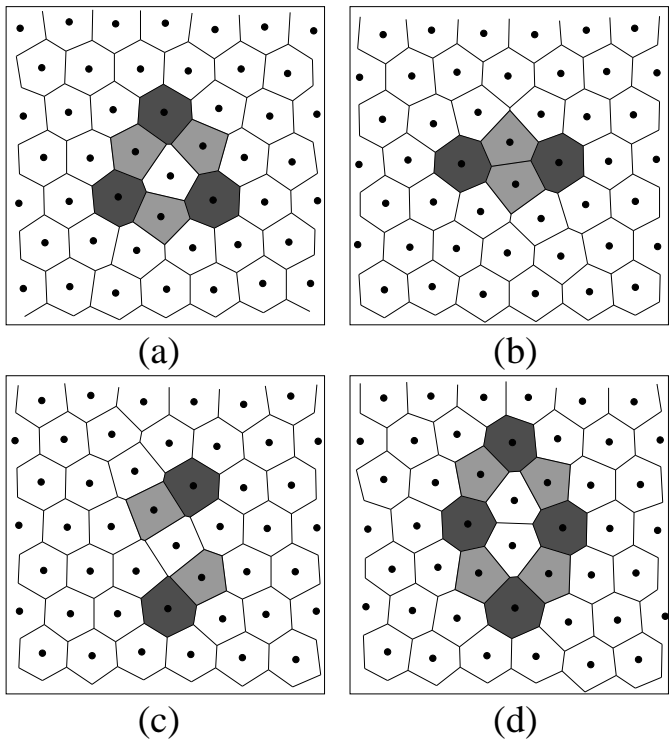}
\caption{Voronoi cell construction for commonly observed
interstitial defect configurations. Colloid positions
are indicated by dots.  The Voronoi cells are colored according to the
coordination number $z_i$ of the colloids: white, $z_i=6$; light grey, 
$z_i=7$; dark grey, $z_i=5$.
Only a $6a_0 \times 6a_0$ portion of the full system is shown.
(a) A threefold symmetric interstitial, $I_3$. 
(b) A twofold symmetric interstitial, $I_2$. 
(c) A disjoint twofold symmetric interstitial, $I_{2d}$.
(d) A fourfold symmetric excited configuration $I^\prime_4$.}
\label{fig:defecttypes2}
\end{figure}

Figure \ref{fig:defecttypes2} shows
the main configurations for the interstitial defect.
Fig.~\ref{fig:defecttypes2}(a) illustrates a
threefold symmetric interstitial, $I_3$, composed of
a triangular arrangement of $z_i=5$ and $z_i=7$ colloids
centered around the interstitial.
A twofold symmetric interstitial configuration, $I_2$, appears
in Fig.~\ref{fig:defecttypes2}(b).
This configuration can split to become a disjoint twofold
symmetric interstitial, $I_{2d}$, shown in Fig.~\ref{fig:defecttypes2}(c).
The interstitial defect forms an excited configuration that is similar in form
to that seen for the vacancy defect.  Termed a fourfold symmetric
excited configuration, 
$I^\prime_4$, this interstitial configuration is shown in
Fig.~\ref{fig:defecttypes2}(d).
As in the vacancy case, this excited configuration 
persists only for short times, but as we describe in Section V.B,
it plays an important role in the
motion of the defect.

\section{Defect motion and Local Burgers Vector Directions}

To study the motion of the two types of defects, we must identify the
location of each defect.  We define the defect position ${\bf r}_d$
to be at the center of the $N_{z_i \ne 6}$ colloids with 
$z_i \ne 6$ that are present in the system, 
\begin{equation}
{\bf r}_d=\frac{1}{N_{z_i \ne 6}}\sum_{i=1}^N{\bf r}_i(1-\delta(z_i-6)) .
\end{equation}
With this measure, we can follow the trajectories of a defect under
thermal diffusion.

\begin{figure}
\includegraphics[width=3.5in]{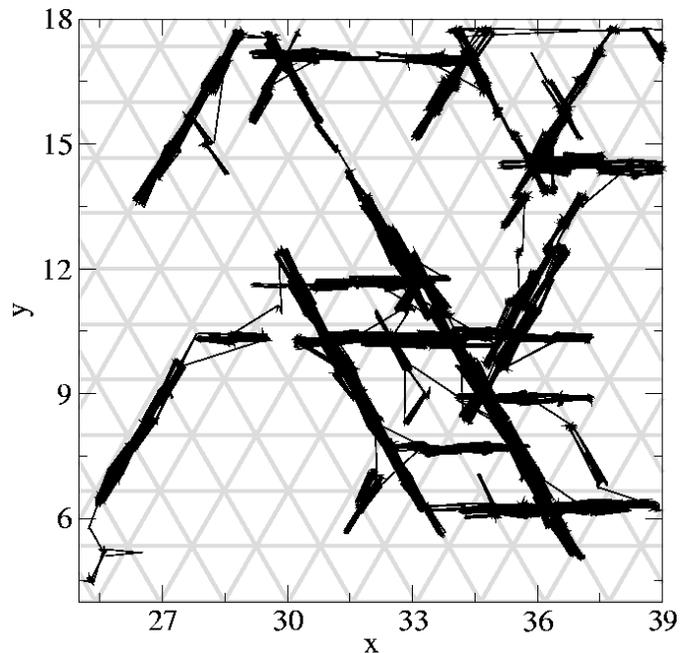}
\caption{Diffusion of a vacancy defect at $f_T/f_T^m=0.4$.  
Only a $14a_0 \times 14a_0$ portion of the
full system is shown.
Dark lines: trajectory followed by the vacancy 
over a period of 
$4.3 \times 10^7$
time steps.
Light lines: long-time average 
location of the crystalline lattice through which the vacancy
is moving.  Over short times, the vacancy diffuses in the directions of
the major axes of the crystal.
}
\label{fig:crystallattice}
\end{figure}

\subsection{Isotropic temperature}

In Fig.~\ref{fig:crystallattice} 
we illustrate the trajectory of a vacancy defect over a period of 
$t=4.3 \times 10^7$
simulation time steps at a temperature of 
$f_T/f_T^m = 0.4$.   
We also indicate the time-averaged location of the background triangular 
lattice to show that the defect diffusion follows the main crystalline 
lattice directions, $[1 0]$, $[0 1]$ and $[\overline{1} 1]$.
The defect diffuses by gliding along the lattice directions in 
a one-dimensional random walk, punctuated by occasional reorientation 
transitions. 
This suggests that at short times the diffusive motion
has one-dimensional characteristics, 
while for long times the diffusion is isotropic. 
The trajectories we observe are very similar to those found in the
vacancy experiments of Ref.~\cite{Pertsinidis} 
where the defect moved along the symmetry directions of the crystal. 
We find that interstitial defects move in a similar manner.
The reorientation transitions that occur during diffusion are mediated
by the formation of either a threefold symmetric configuration or
one of the excited configurations illustrated in Figs.~\ref{fig:defecttypes}
and \ref{fig:defecttypes2}.  We describe the reorientations in more
detail below.

In order to characterize the motion along the different crystalline axis, 
we require a definition of the orientation of the defect as a function of
time.  The orientation of a dislocation is readily obtained using the
Burgers vector ${\bf b}$; however, a point defect has a net
Burgers vector of ${\bf b}=0.$  We can construct local Burgers vectors
${\bf b}_j^l$
by associating each of the $N_{z_i=5}$ colloids that have $z_i=5$ 
with the closest $z_i>6$ colloid to
form a dislocation, such that each $z_i=7$ colloid is paired with only
one $z_i=5$ colloid, and under the constraint that the total length, 
$\sum_{j=1}^{N_{z_i=5}}|{\bf b}_j^l|$, is
minimized.

\begin{figure}
\includegraphics[width=3.5in]{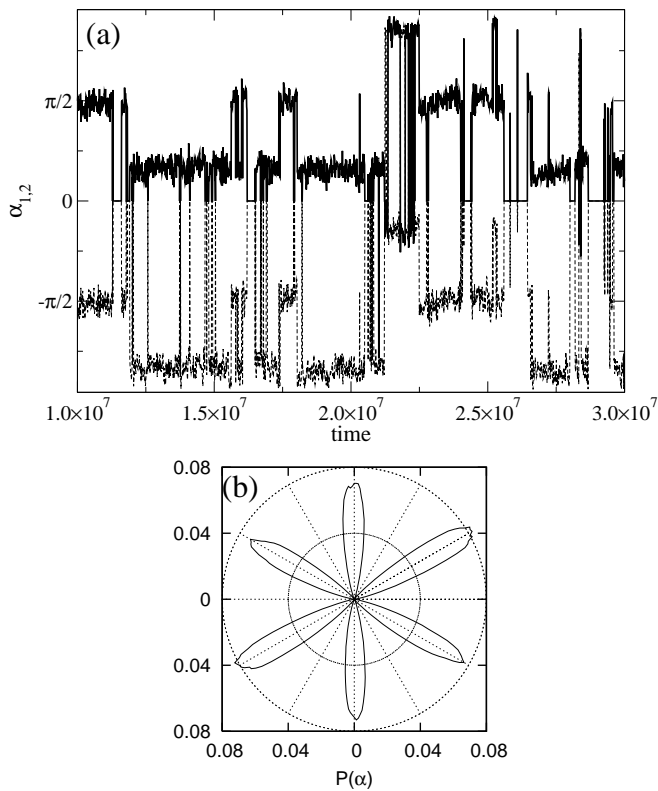}
\caption{
(a) Time series of the orientations 
$\alpha_1$ (upper dark line) and $\alpha_2$ (lower dashed line)
determined from the two local Burgers vectors
${\bf b}_1^l$ and ${\bf b}_2^l$
in the gliding vacancy defect configurations $SV$ and $V_{2a}$
for a system with $f_T/f_T^m=0.4$.
(b) From the same system, a polar plot of the
probability $P(\alpha)$ of observing a vacancy defect
oriented at an angle $\alpha$ with respect to the zero $x$ axis.
}
\label{fig:burgersangles}
\end{figure}

The twofold symmetric configurations, $V_{2a}$, $I_2$, and $I_{2d}$, along with
the split vacancy configuration $SV$, all have only two local Burgers vectors
${\bf b}_j^l$ which are close to being parallel to each other.
For these configurations, which we refer to as gliding configurations, 
we define
the orientation of the defect to be in the direction 
perpendicular to one of the local
Burgers vectors.
We also identify the angle $\alpha_j$ which each 
Burgers vector makes with the zero $y$ axis.
In Fig.~\ref{fig:burgersangles}(a)
we plot the time series of $\alpha_1$ and $\alpha_2$ for a system with a
vacancy defect diffusing at 
$f_T/f_T^m=0.4$.  
The angles are only defined during
the time periods when the vacancy is in the configuration $V_{2a}$ or
$SV$, but the vacancy spends most of the time in one of these two 
configurations,
which have $|\alpha_1-\alpha_2| \simeq \pi$.  Figure \ref{fig:burgersangles}(a)
shows that $\alpha_1$ and $\alpha_2$ remain fixed at a particular angle for
extended periods of time, punctuated by relatively rapid changes to a
new angle.  A histogram of the $\alpha$ values with both $\alpha_1$ and
$\alpha_2$ combined, plotted in polar coordinates in
Fig.~\ref{fig:burgersangles}(b), indicates that $\alpha$ 
is correlated with the
six directions of the crystal lattice.  Within our sampling error, all six
of the angles appear with equal probability.

\subsection{Anisotropic temperature}

Many two-dimensional systems contain some type of anisotropy which
could originate in the particle-particle interactions or from an underlying
weak periodic substrate modulation \cite{Chowdhury,Wei,Radzihovsky}. 
Since we found that the defects move along the symmetry directions 
of the crystalline lattice, it may be
expected that if some form of anisotropy is added, the motion 
may be more prominent along certain directions. 
A similar system has recently been realized experimentally 
in a colloidal system composed of superparamagnetic particles
\cite{Maret2}. With the
application of an external magnetic field, 
the interactions between the colloids could be made anisotropic.
As a result, dislocations formed in the system with a preferred orientation,
resulting in melting along the direction of the applied magnetic field. 

\begin{figure}
\includegraphics[width=3.5in]{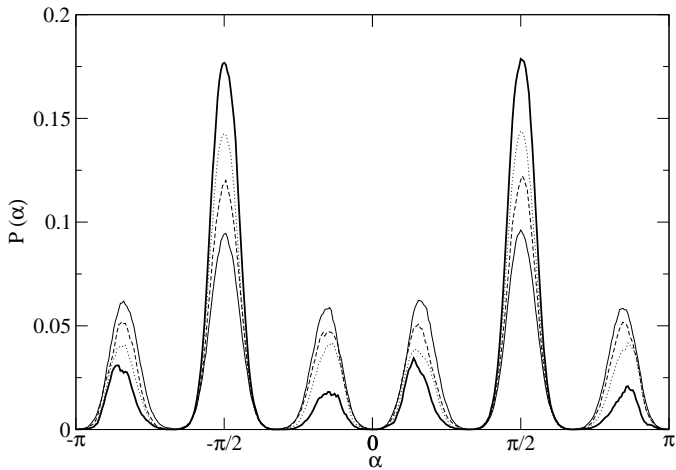}
\caption{$P(\alpha)$, the histogram of $\alpha$ values observed during
a simulation run, for different temperature anisotropies
in a system with a vacancy defect.
In all cases $f_T^y/f_T^m=0.62$. 
Thick continuous line: $f_T^x/f_T^y=0.1$,
dotted line: $f_T^x/f_T^y=0.3$,
dashed line: $f_T^x/f_T^y=0.5$,
and thin continuous line: $f_T^x/f_T^y=0.7$.
}
\label{fig:anisotropic}
\end{figure}

We consider the effect of adding an anisotropic temperature to our system
by setting $f_T^x/f_T^y < 1$, where $f_T^y$ is fixed
at 
$f_T^y/f_T^m=0.62$.  
Under these conditions, the symmetric
diffusion disappears and the defects move preferentially in one direction.
In Fig.~\ref{fig:anisotropic} we plot $P(\alpha)$ for several different
temperature anisotropy ratios.
As $f_T^x/f_T^y$ is decreased from 1, the value of $P(|\alpha=\pi/2|)$
increases dramatically while $P(\alpha)$ for the other four crystal
directions decreases.  This indicates that the defect spends most of the time
with $\alpha$ oriented along the $y$ direction, meaning that the local
Burgers vector is oriented along the $x$ direction.  Reorientation
transitions for the defect occur much less frequently than in the 
isotropic case, causing the defect to undergo one-dimensional diffusion
for much longer periods of time, and resulting in a highly anisotropic
diffusion over time.  Such anisotropic diffusion is consistent with the
results obtained in the experiment of Ref.~\cite{Maret2}.

\section{Properties of the defect diffusion}

We have shown that the defect trajectory follows the main crystalline 
directions.  The mechanism of defect motion is primarily a gliding
process.  In the case of a vacancy defect, the vacancy can glide
in the configurations $SV$ or $V_{2a}$, which have parallel local
Burgers vectors and a well-defined glide direction.
The one-dimensional 
gliding motion is interspersed with direction switching transitions 
that occur by means of the $V_3$ configuration or the excited 
$V^\prime_4$ configuration.
Both of these configurations have higher symmetry than the gliding 
configurations 
but also
have nonparallel local Burgers vectors, and hence no well-defined glide
direction.
Thus, when the vacancy enters a transition configuration, the symmetry breaking
that defined a glide direction is lost.  The vacancy spends a relatively
short time in the transition configuration before reentering one of the glide
configurations, which may have the same or a different glide orientation
than the glide configuration 
which preceded the transition configuration.  The high
symmetry of the transition configurations permit the defect to reorient its
direction of motion along a new lattice direction.
Interstitial defects undergo similar sequences of gliding and reorientation,
with the $I_2$ and $I_{2d}$ configurations 
providing the gliding motion, and the
$I_3$ and $I^\prime_4$ configurations permitting the reorientation.

The vacancy position remains stationary when entering a transition 
configuration such as $V_3$, but the position of the interstitial shifts
slightly when switching between the $I_2$ or $I_{2d}$ and $I_3$ 
configurations, providing an additional mechanism for interstitial motion
that does not exist for the vacancy defect.  
It is also possible for interstitial motion to
occur through a `breathing' process caused by transitions between
the $I_3$ and $I^\prime_4$ configurations.

\begin{figure}
\includegraphics[width=3.5in]{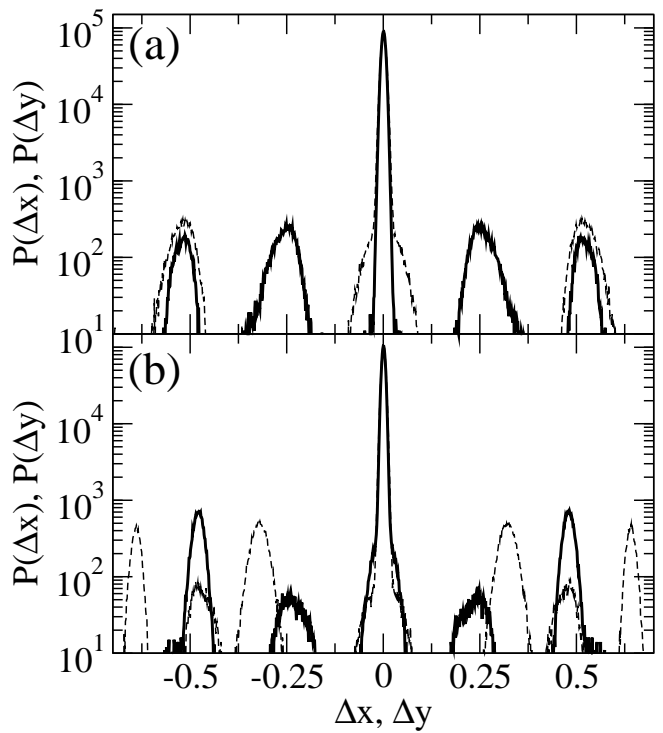}
\caption{(a)
 Histograms of the discrete jumps $\Delta x$ (solid line) and
$\Delta y$ (dashed line) for a vacancy defect.
(b) 
 Histograms of the discrete jumps $\Delta x$ (solid line) and
$\Delta y$ (dashed line) for an interstitial defect.
In this figure, the jump distances are measured in their 
natural units. $\Delta x$ is given in units of $a_0$, 
and $\Delta y$ is given in units of $\sqrt{3}a_0/2$}
\label{fig:histojump}
\end{figure}

We highlight the difference in the mobility mechanisms 
for the two types of defects  
by measuring the individual jumps of the defects from one lattice
position to the next.  We define
$\Delta x_i=({\bf r}_{d,i}(t+dt)-{\bf r}_{d,i}(t)) \cdot {\bf {\hat x}}$ and
$\Delta y_i=(2/\sqrt{3})({\bf r}_{d,i}(t+dt)-{\bf r}_{d,i}(t)) 
\cdot {\bf {\hat y}}$, 
where we take $dt=200$ simulation time steps.
Histograms of both $\Delta x$ and $\Delta y$ are plotted in
Fig.~\ref{fig:histojump}(a) for a vacancy defect and in
Fig.~\ref{fig:histojump}(b) for an interstitial defect.
The vacancy defect moves only by a glide mechanism. 
The process of switching between $V_{2a}$ and $SV$ translates
the defect center by half a lattice constant along one of the 
main crystal axis directions. 
This produces four non-zero values of $\Delta x$, $\pm \cos(0)/2$ and 
$\pm \cos(\pi/3)$, and two non-zero values for
$\Delta y$, $\pm \sin(\pi/3)/\sqrt{3}$. 
Since we are plotting the histogram of both 
$\Delta x$ and $\Delta y$ in their natural units, 
which are $a_0$ and $\sqrt{3}a_0/2$ 
respectively, we observe peaks 
at $\Delta x=\pm 0.25$ and $\pm 0.5$ and at $\Delta y=\pm 0.5$
in Fig.~\ref{fig:histojump}(a).
The same peaks appear for the interstitial defect in 
Fig.~\ref{fig:histojump}(b) since the gliding mechanism also operates
in this case, but there are now additional peaks in $\Delta x$ and
$\Delta y$.
These peaks are produced by the other two processes that can move 
the interstitial defect: 
the switch between the $I_2$ or $I_{2d}$ and $I_3$ configurations, and the
breathing motion between the $I_3$ and $I^\prime_4$ configurations.
These two processes move the defect center half a lattice constant along 
directions that lie between the main crystalline directions, at
$\pi/6$, $\pi/2$, $5\pi/6$, $7\pi/6$, $3\pi/2$, and $11\pi/6$. 
The result is peaks at
$\Delta x = \pm \cos(\pi/6)/2 \simeq \pm 0.43$ and 
peaks at $\Delta y = \pm \sin(\pi/6)/\sqrt{3} \simeq \pm 0.29$ 
and $\Delta y = \pm 2\sin(\pi/6)/\sqrt{3} \simeq \pm 0.58$. 

\subsection{Differences in defect diffusion for vacancies and interstitials}

\begin{figure}
\includegraphics[width=3.5in]{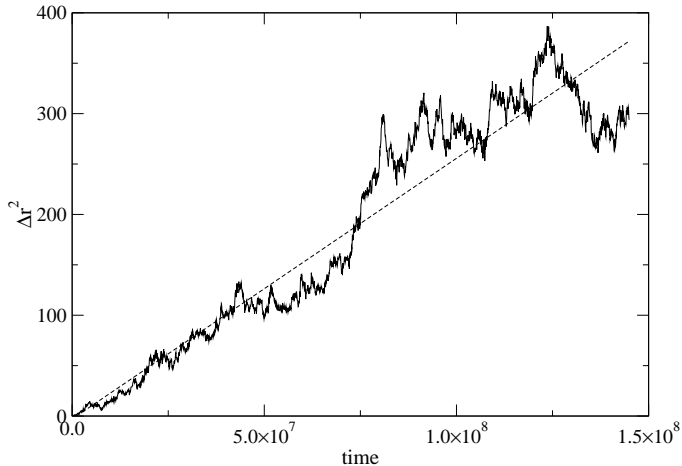}
\caption{
$\Delta r^2$ versus time showing the
diffusion of the vacancy defect at $f_T/f_T^m=0.62$.  The dashed line is
a linear fit, consistent with linear two-dimensional diffusion.
}
\label{fig:diffuse1d}
\end{figure}

\begin{figure}
\includegraphics[width=3.5in]{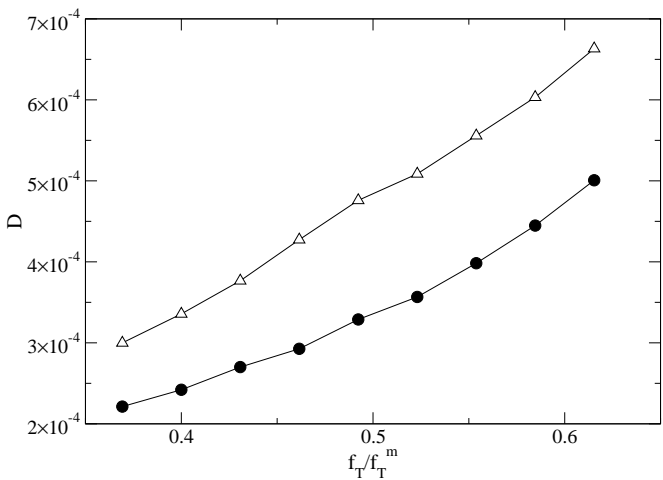}
\caption{
Diffusion constant $D$ as a function of temperature $f_T/f_T^m$.
Filled circles: vacancy defect.  
Open triangles: interstitial defect.}
\label{fig:diffconstant}
\end{figure}

For short times, the defect trajectory follows the main crystalline
directions, as was shown in Fig.~\ref{fig:crystallattice}.  On the
longer time scales, however, we find a linear diffusive behavior when
we measure the distance that the defect has traveled from its original
location as a function of time, $\Delta r^2 = |{\bf r}_d(t)-{\bf r}_d(0)|^2$.
This quantity is plotted in Fig.~\ref{fig:diffuse1d} for a vacancy
defect at $f_T/f_T^m=0.4$.  We find that $\Delta r^2$ increases
linearly with time, as indicated by the fit in the figure.  Linear
diffusion of monovacancies was also observed in the
experiments of Ref.~\cite{Pertsinidis}.

To compare the mobility of interstitial and vacancy defect, we
measure the diffusion constant $D$, given by
\begin{equation}
D=\left\langle \frac{|{\bf r}_d(t+dt)-{\bf r}_d(t)|}{dt}\right\rangle ,
\end{equation}
with $dt=200$ simulation time steps,
at a series of temperatures for each defect type.
The results are plotted in 
Fig.~\ref{fig:diffconstant}.
The diffusion constant increases with temperature in each case. 
Importantly, {\it the interstitial
defect diffuses significantly faster than the vacancy defect} at all
temperatures.

\subsection{Probability of different defect configurations}

\begin{figure}
\includegraphics[width=3.5in]{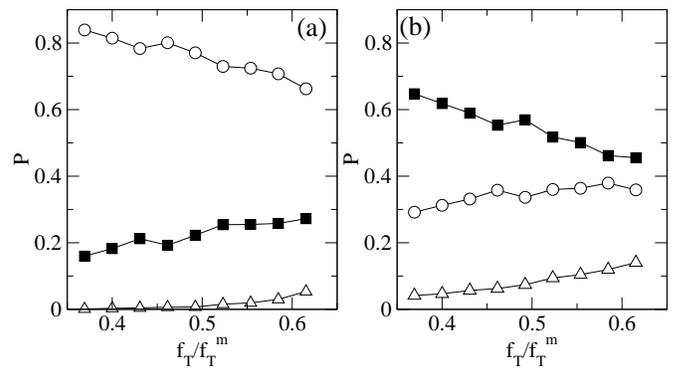}
\caption{
(a) 
Probability to observe different defect configurations for the vacancy
defect.  Open circles: $P(SV)+P(V_{2a})$; 
filled squares: $P(V_3)$; open triangles:
$P(V^\prime_4)$.
(b) 
Probability to observe different defect configurations for the interstitial
defect.  Filled squares: $P(I_3)$; open circles: $P(I_2)+P(I_{2d})$; 
open triangles:
$P(I^\prime_4)$.
}
\label{fig:probabstates}
\end{figure}

To help account for the faster diffusion of the interstitial defect
compared to the vacancy defect,
we compare the time the defects spend in the 
different topological configurations.
In Fig.~\ref{fig:probabstates}(a) we plot the probability $P$ of observing
each configuration as a function of temperature $f_T/f_T^m$ for the vacancy
defect, and compare this to Fig.~\ref{fig:probabstates}(b) which shows the
probability of observing interstitial defect configurations versus temperature.

Figure \ref{fig:probabstates}(a) shows
that the vacancy defect spends most of the time in the split ($SV$) 
and the twofold crushed ($V_{2a}$) 
configurations, which both allow for gliding
motion.  The excited configuration $V^\prime_4$ occurs very infrequently 
and does not contribute significantly to the diffusive process.
As the temperature increases, the transition states $V_3$ and $V^\prime_4$
occur with greater probability, indicating that the defect can reorient
more easily.
A greater number of defect reorientations combined with higher
thermal excitation leads to a higher diffusion 
constant at higher temperature.  Since the gliding motion is one-dimensional,
the defect can only move away from its starting position as rapidly as
a one-dimensional random walk if no reorientation occurs.
Reorientation of the glide direction produces a more
two-dimensional diffusion which allows the defect to move away
more rapidly from its starting position.

The interstitial defect spends much more time in the transition configuration
$I_3$ than in any other 
configuration, as indicated in Fig.~\ref{fig:probabstates}(b).
The excited $I^\prime_4$ configuration 
for the interstitial defect also appears with much higher 
probability than the excited $V^\prime_4$ configuration for the vacancy
defect case.  Overall, this produces a much higher probability for
reorientation of the direction of motion of the interstitial defect, and
makes its diffusion {\it more two-dimensional} than that of the vacancy
defect at all temperatures considered here.  
The result is a greater mobility of the
interstitial with respect to the vacancy.

\section{Discussion}
An open question is how general our results are for defect motion in 
other two-dimensional systems 
where the particles have different types of interaction potentials.
Although it is beyond the scope of this paper to address this       
issue in detail, 
we believe that our results should hold for any two-dimensional system
with relatively soft particle-particle interactions in which the ground
state is a triangular lattice.
Very short range potentials, such as hard sphere interactions,
are likely to produce significantly different behavior.
It is possible for two-dimensional systems to have some other type of
ordered ground state, such as a square or rectangular
lattice.
These lattices have different available gliding directions compared to
the triangular lattice, so there could be either fewer or more modes of
motion available to the defects.
This would be an interesting issue to explore in the future.  

Our results could have some implications for 
the mobility of defects in other two-dimensional systems where the 
number of vacancies or interstitials
can be carefully controlled, such as superconducting vortices 
\cite{Harada,Reichhardt} or colloids \cite{Bechinger} 
interacting with periodic substrates. At particle
densities where there is one particle per substrate minimum, termed
a matching density,
the system is free of topological
defects, while vacancies appear for slightly lower particle densities and
interstitial defects form at slightly higher particle densities.
If the interstitial defects remain more mobile than vacancy defects in
the presence of a substrate,
the interstitials should depin more readily than the
vacancy defects. 
In the case of superconducting vortices, this would imply that the critical
current drops off more rapidly above the matching density (when interstitials
are present) than below the matching density (when vacancies are present).
This is agreement with earlier observations \cite{Reichhardt}.

Vacancy and interstitial defect motion has also been studied in
three-dimensional systems in the context of atomic crystals.  Such 
systems are not only massive, making them sensitive to vibrational modes,
but also can form a variety of different lattice structures depending
on the details of the atomic interactions.  A convenient method for
producing isolated vacancies or interstitials in an atomic crystal is
through ion damage processes.  We note that in this context, 
unusually high mobility of self-interstitial atoms (SIAs) has been observed and
attributed to the high likelihood of rotational motions for
the dumbbell atomic configurations associated with the SIA
\cite{Schilling}.  Although the
dimensionality of the resulting motion is different, this resembles our
observation that interstitial defects are more mobile than vacancy
defects due to the greater likelihood of defect reorientation.

Another issue is the relative stability of interstitial and vacancy
defects.
Due to the geometry of our system, which has periodic boundary conditions,
both the interstitial and vacancy defects are completely stable since there
is no free surface where the defects can annihilate.
In a real crystal, the more mobile species of defect will reach
the edge more rapidly and annihilate.
Thus, one implication of the faster mobility of interstitial defects
is that a crystal that forms with an initial population of vacancy and
interstitial defects will, over time, experience a greater loss of
interstitial defects to the crystal edge compared to vacancies,  
particularly if there is a strain field that helps drive the defects
to the edge.  Both
interstitials and vacancies can still annihilate within the crystal, but
the difference in edge annihilation could result in a dominant population
of the slower moving vacancy defects in the relaxed crystal.

\section{Conclusions}

In summary, we have studied the topological configurations and dynamics 
of individual vacancy and interstitial defects in a triangular 
two dimensional colloidal crystal. 
We use a Voronoi cell construction to characterize the different topological
configurations assumed by the defects.
For vacancy defects, we find the same configurations that were observed 
in recent experiments. 
We show that interstitial defects, which were not studied in the experiment,
appear in distinct configurations and 
have a significantly higher mobility than the vacancy defects.
The mobility is affected by short lived excited configurations which 
have four or more fivefold or sevenfold coordinated colloids surrounding
the defect.
The defect diffusion process 
is one dimensional at short time scales when the defects glide along the
symmetry directions of the crystal.  Periodically, the defects enter a
transition configuration which has higher symmetry than the glide
configuration.  This permits the defect to reorient its glide direction
when it returns to a glide configuration after a short period of time.
As a result of these reorientations, linear two-dimensional diffusion occurs
in the long time limit.
Application of an anisotropic temperature causes the diffusion process 
to favor one of the permitted gliding directions. 
We describe the mechanisms of defect translation in detail,
and show that the vacancy defect has one mode of motion, but the interstitial
defect has three.  It is these additional
modes of motion that give the interstitial defect a has a much higher
probability of undergoing a reorientation transition than the vacancy
defect at a given temperature, causing the interstitial diffusion to be
more two-dimensional in character and therefore faster.

\section{Acknowledgments}

We thank Z. Nussinov, A.F. Voter, and G. Zim\'anyi for useful discussions. 
This work was carried out under the auspices of the National Nuclear
Security Administration of the U.S. Department of Energy at Los Alamos
National Laboratory under Contract No.~DE-AC52-06NA25396.

\end{document}